\documentclass[namedreferences]{solarphysics}
\usepackage[optionalrh,solaromanenum]{spr-sola-addons} 
\usepackage{epsfig}
\usepackage{epstopdf}
\usepackage{graphicx}                    
\usepackage{amssymb}                    
\usepackage{iondefs}
\usepackage{upgreek}

\usepackage{color}                       
\usepackage{url}                         

\begin{document}

\begin{article}

\begin{opening}

\title{Performance Testing of an Off-Limb Solar Adaptive Optics System}

%
\author{G.~E.~\surname{Taylor}$^{1}$\sep
        D.~\surname{Schmidt}$^{2}$\sep
        J.~\surname{Marino}$^{2}$\sep
        T.~R.~\surname{Rimmele}$^{2}$\sep
        R.~T.~J.~\surname{McAteer}$^{1}$\sep
       }

%

%
  \institute{$^{1}$ New Mexico State University
                     email: \url{seryddwr@nmsu.edu; mcateer@nmsu.edu}\\ 
                     $^{2}$ National Solar Observatory
                     email: \url{dschmidt@nso.edu; marinoj@nso.edu; rimmele@nso.edu}
             }

\begin{abstract}
Long-exposure spectro-polarimetry in the near-infrared is a preferred method to 
measure the magnetic field and other physical properties of solar prominences. 
In the past, it has been very difficult to observe prominences in this way with 
sufficient spatial resolution to fully understand their dynamical properties. 
Solar prominences contain highly transient structures, visible only at small 
spatial scales; hence they must be observed at sub-arcsecond resolution, with 
a high temporal cadence. An adaptive optics (AO) 
system capable of directly locking-on to prominence structure away from the 
solar limb has the potential to allow for diffraction-limited 
spectro-polarimetry of solar prominences. In this paper, 
the performance of the off-limb AO system and its expected performance, at the 
desired science wavelength {\CaII} 8542 {\Ang}, are shown.
\end{abstract}

%

\end{opening}

%

\section{Introduction}
\label{sec:intro}  
Solar prominences consist of relatively cool, dense plasma, which is suspended
above the solar surface. Filaments and prominences represent the same 
phenomenon, 
viewed differently \cite{TH,2010SSRv..151..243L,2010SSRv..151..333M}.
The prominence itself forms within or directly beneath a magnetic
flux rope \cite{2010SSRv..151..333M,2014IAUS..300...15B}. 
A flux rope is a long, twisted magnetic field region, 
which is nearly force-free \cite{TH,2010SSRv..151..333M}. Such a flux
rope may form either from the shearing of magnetic arcades, or it may emerge
from below the photosphere, already formed \cite{2010SSRv..151..333M}. 
One main reason that the study of solar prominence should be considered
important is that they are often involved in coronal mass ejections (CMEs).
CMEs occur when a flux rope becomes unstable, {\it e.g.} when there is emerging 
magnetic flux, coming from below \cite{2011LRSP....8....1C}. Since flux 
ropes very often contain solar prominences, prominences may show 
instabilities that could be used to predict an 
imminent CME \cite{2011LRSP....8....1C}.

Despite continuing study of solar prominences, there is still much that is 
unknown about their physical and magnetic structure at small spatial 
scales \cite{2010SSRv..151..243L,2010SSRv..151..333M,2014IAUS..300...15B}. 
There are ambiguities related to the broadening of spectral lines 
of prominences that are attributed to unresolved fine 
structure \cite{2010SSRv..151..243L}. Thus the ability to measure 
spectra of solar prominences at very fine spatial scales is necessary 
to the understanding of solar prominences.
Since solar prominence plasma interacts 
with the magnetic field in which it resides, an understanding of 
prominence dynamics can only be understood by the inversion of 
spectro-polarimetric data \cite{2010SSRv..151..333M}. 
There are many ambiguities about prominence 
behavior, with respect to their magnetic fields, at the small spatial scales 
because these structures are unresolved \cite{2010SSRv..151..333M}.
Space-based instruments are capable of 
imaging solar prominences at very fine spatial 
scales \cite{2011Natur.472..197B}. 
However, to the knowledge of the authors, none has 
been launched that can perform spectro-polarimetry on a solar prominence, 
which is necessary for the understanding of prominence magnetic 
fields \cite{TH,2010SSRv..151..333M}. 

Ground-based telescopes have instruments which are capable of 
taking spectral and spectro-polarimetric data, 
for example the Interferometric BIdimensional Spectrometer 
(IBIS; \opencite{2006SoPh..236..415C}) and the Facility 
Infrared Spectrometer (FIRS;\opencite{2010MmSAI..81..763J}) on the {\it Dunn 
Solar Telescope} (DST) in New Mexico.
Current solar adaptive optics (AO) systems are designed to utilize broad-band 
light and are thus confined to locking onto structures on the disk of the Sun; 
prominences are invisible in broadband light \cite{2011LRSP....8....2R}. 
This limits 
the usefulness of current solar AO systems. They can only be used to 
correct prominence images, and hence provide high resolution spectral or 
spectro-polarimetric 
data when there is a pore or other dark feature directly adjacent to the 
part of the limb near the prominence, as was done by 
\inlinecite{2013hsa7.conf..786O}. These data, however, are fundamentally 
limited in resolution because AO systems can only provide their best 
correction extremely close to the point upon which they are locked 
\cite{har_98}.
Only a purpose-built AO system that can directly lock onto 
solar prominence structure can allow for spectroscopic and spectro-polarimetric 
data at the diffraction limit of the telescope; thus allowing for an 
increase in humankind's understanding of solar prominences.

The authors are the first group to construct an off-limb solar AO system 
mainly because of the difficulty in measuring the incoming wavefront using 
only light from 
solar prominences. The main issue is photon flux. {\Ha} light from solar 
prominences was chosen for wavefront sensor (WFS) of the off-limb solar AO 
system because prominences 
emit very brightly at this wavelength, relative to other spectral lines.
Even so, the required 0.5 - 0.7 {\Ang} filter bandwidth transmits very 
little flux. 
Thus, it is necessary to use all available {\Ha} light for the WFS.
An entire AO system was designed and optimized around this WFS.
The difficulty of measuring the wavefront, coupled with the need to design an
entirely new AO system, is the reason why the authors are the first to build an 
off-limb solar AO system. (For a review of specific 
problems of solar AO in general, see the earlier, comprehensive reviews 
of \inlinecite{1998SPIE.3353...72R} and \inlinecite{2011LRSP....8....2R}.)

A correlating Shack-Hartmann wavefront sensor 
(SHWFS) was chosen for the off-limb solar AO system, due to its well known 
characteristics, though 
it is not the only type of WFS that was considered \cite{2012ASPC..463..321T}.
A standard SHWFS uses point source images directly, {\it i.e.} from 
a star or laser, to measure the wavefront. It does this by 
breaking the beam of light into small pieces, each representing 
a portion of the telescope aperture. The position of each star image 
directly corresponds to the wavefront slope at each position in the 
telescope aperture. A map of the wavefront is created from these 
slope measurements \cite{har_98}; see Figure \ref{pupil}.
Correlating SHWFS have been used in solar AO for several years
\cite{1998SPIE.3353...72R,2011LRSP....8....2R}.
The difference between a correlating SHWFS and a conventional one is that 
each sub-aperture image is cross-correlated with a reference image, typically 
one sub-aperture image from a single frame. The maximum of each 
cross-correlation function shifts with the image motion, and thus the 
changing wavefront slope 
in each of the SHWFS 
sub-apertures \cite{1998SPIE.3353...72R,2011LRSP....8....2R}.

In this paper the results of some of the early tests of 
a closed-loop AO system on the DST are shown, 
based on a correlating SHWFS. In Section \ref{sec:pred},
 \inlinecite{2013SPIE.8862E..0CT} is reviewed, which details the performance 
which was expected to be achieved from the off-limb solar AO system. In 
Section \ref{sec:setup}, the setup of the experiment which was used to verify 
the performance predictions is detailed. In Section \ref{sec:res}, the results 
of these experiment are enumerated and it is shown how they relate to the 
predictions in Section \ref{sec:pred}. 

\section{Predicted Performance}\label{sec:pred}

\begin{figure}[t]
\centering
\includegraphics[width=0.8\textwidth]{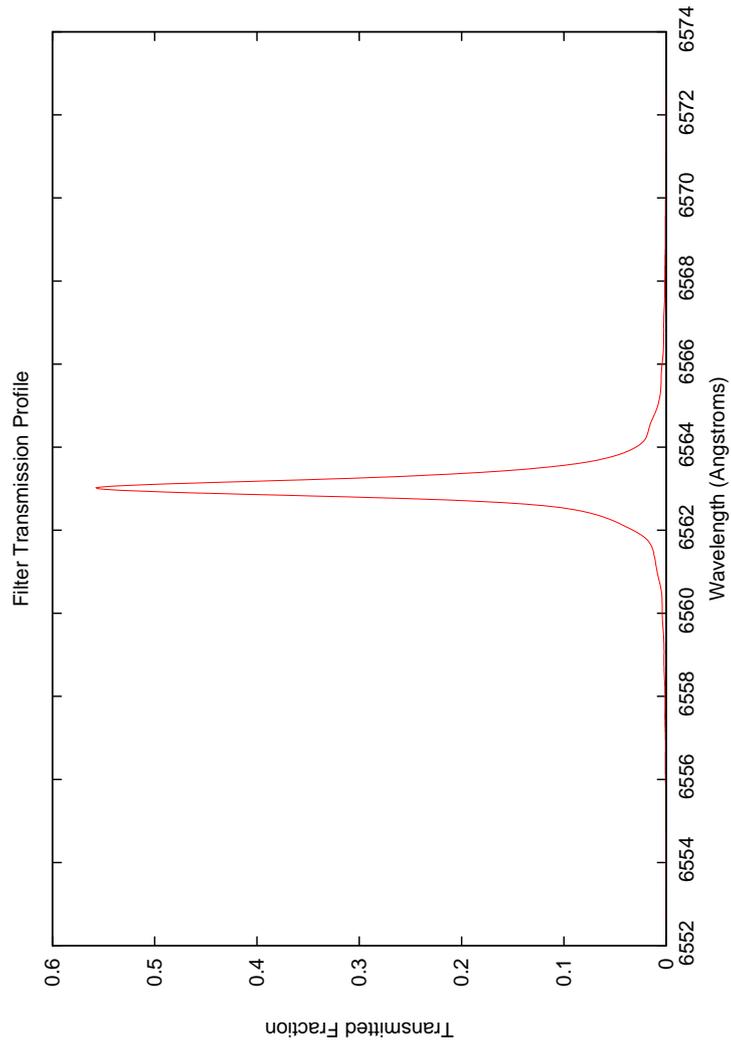}
\caption{Filter transmission profile of the custom {\Ha} filter.}
\label{fig:filter}
\end{figure}

As was stated in Section \ref{sec:intro} there is very little light 
available for the SHWFS.
Indeed, there is so little flux available, even in {\Ha} light, a custom made 
{\Ha} filter with very high transmission was used; see 
Figure \ref{fig:filter}. This same type of filter will be used on the 
Visible Broadband Imager on the {\it Daniel K. Inoue Solar Telescope} 
(DKIST, formerly ATST; \opencite{2014IAUS..300..362R}).
A Dalsa Falcon VGA300 HG machine-vision camera was selected, due to its high 
sensitivity, moderate read-noise, high speed, and low cost, coupled with an 
in-house lenslet array, to make the SHWFS. To achieve the required 
speed, only a small portion of the camera chip ($160 \times 160$ pixels) 
was read-out.

Using the high transmission filter and sensitive camera, 
the SHWFS is still photon starved; thus 
it is necessary to make trade-offs in design parameters. 
If the wavefront were finely sampled, by dividing the SHWFS field of 
view into many sub-apertures, a great deal of light would be lost and 
the SHWFS would no longer be able to sense the wavefront at a sufficiently 
high frame-rate to allow for AO correction. 
Therefore, fewer sub-apertures 
were used for this SHWFS than for the SHWFS used for on-disk observation at the 
DST \cite{1998SPIE.3353...72R}. 
A large field of view (FOV) per sub-aperture is also needed, to allow for the 
tracking of solar prominence features that are large in comparison to features 
seen on the disk \cite{2012ASPC..463..321T}. Finally, the 
final pixel scale of the sub-aperture images must be chosen to 
balance resolution and image brightness \cite{1993rtpf.conf..124M}.
The number of sub-apertures used, their FOV, and the final 
pixel scale can be determined analytically, as well as directly measured
\cite{2003SPIE.4853..370S}. 
To verify 
that an off-limb solar AO system is possible and practical, the performance 
of a particularly promising SHWFS configuration was modeled in detail.

\subsection{WFS Noise from Indirect Methods}\label{sec:ind}
It is necessary to 
measure the noise created by the SHWFS 
and to estimate noise from the various other 
sources that are present in a finished AO system, in 
order to predict the ultimate performance of an off-limb solar AO 
system \cite{har_98}.

\begin{figure}[t]
\centering
\includegraphics[width=6cm,clip]{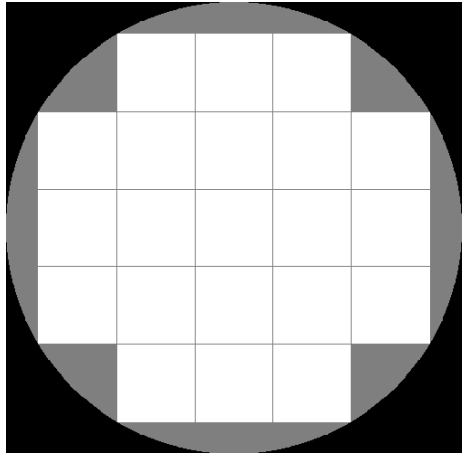}
\caption[Pupils]{The pupil layout for the $5 \times 5$ grid. The outer 
circle is the telescope pupil, 76.2cm in diameter.}
\label{pupil}
\end{figure}

\begin{figure}[t]
\centering
\includegraphics[width=10cm,clip]{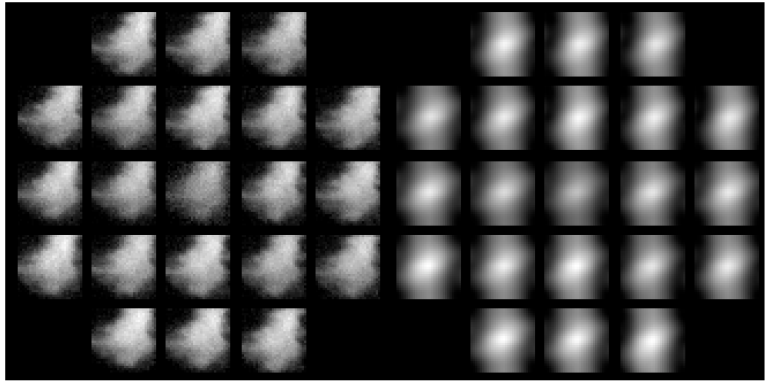}
\caption[Pupils]{SHWFS images, using the Dalsa Falcon VGA300 HG camera. On the 
left are prominence images which have been dark-subtracted and flat-fielded. 
On the right are the cross-correlation peaks.}
\label{WFSIMAGE}
\end{figure}

A SHWFS was setup with a $5 \times 5$ array of sub-apertures 
(see Figure \ref{pupil}), 
each with an FOV
of $30''$ and an image scale of $0.80''$ per pixel.
The image scale and FOV were measured using a resolution target. 
This camera was run with a frame rate of $900 \mathrm {Hz}$ 
and exposure times of $900 \mathrm {\upmu s}$. 
The functional form for noise generated by a correlating SHWFS is given by 
\inlinecite{2011LRSP....8....2R}. However this function was 
explicitly stated to be defined when the Nyquist sampling criterion is 
satisfied, which this system cannot satisfy, 
due to low light levels. \inlinecite{1993rtpf.conf..124M} give 
a more complete version of this formula:
\begin{equation}
\sigma^2_x = \frac{5m^2 \sigma^2_{\mathrm {b}}}{n^2_{\mathrm {r}} \sigma^2_{\mathrm {i}}} \frac{(dp)^2}{(f\lambda)^2}
\qquad [\mathrm{waves}^2],
\label{wfsn}
\end{equation}

\noindent where $\sigma^2_x$ is the total noise and $m$ is the 
full-width at half-maximum 
(FWHM) of the cross-correlation (CC) peak, from the sub-aperture images 
 (see Figure \ref{WFSIMAGE}.).
The quantity $\sigma^2_{\mathrm {b}}$ is the detector noise, which is assumed to 
be read-noise plus photon noise. The read noise was measured to be 
approximately 68 electrons per pixel (see \opencite{BB}, for details).
The photon noise was assumed to be Poisson noise: 
$\sigma_{\mathrm{photon}}^2 = g\bar{S}_{\mathrm{ADU}}$, where $g$ is the camera gain 
and $\bar{S}_{\mathrm{ADU}}$ is the average pixel value, in 
analog-to-digital units (ADUs) \cite{BB}. 
The term $n^2_{\mathrm {r}} \sigma^2_{\mathrm {i}}$ is the energy content 
of the image, $d$ is the sub-aperture size, $p$ is the physical size of the 
pixels, $f$ is the effective focal length of the optical system (telescope plus 
bench optics), and $\lambda$ is the wavelength at which the measurements are 
made, 6563 {\Ang}.
Note that
\begin{equation}
\frac{(dp)^2}{(f\lambda)^2} \approx \frac{(d\theta)^2}{\lambda^2},
\label{theta}
\end{equation}

\noindent where $\theta$ is the image scale, in radians per pixel, and $d$ is 
13.07cm (see Figure \ref{pupil}).

To measure $m$, the CC peaks for the $5 \times 5$ 
array were used \cite{2011LRSP....8....2R}.
Following the method of \inlinecite{1993rtpf.conf..124M}, 
a two-dimensional (2D) Gaussian function was fit to the central portion of 
each CC peak. The value of $m$ is the average FWHM of all the 
CC peaks.The detector noise is again denoted as $\sigma_{\mathrm {b}}$. 

The energy of the image is
\begin{equation}
n^2_{\mathrm {r}} \sigma^2_{\mathrm {i}} = \sum_{j}|S_{j} - \bar{S}|^2,
\label{energy}
\end{equation}

\noindent where $S_{j}$ and $\bar{S}$ are the signal per pixel and the 
average pixel value, measured in electrons \cite{1993rtpf.conf..124M}.

The result of Equation (\ref{wfsn}) was calculated, for each of the 2000 
frames, at each exposure time, and then averaged over those 2000 frames.
To make the smooth curves, as shown in Figure \ref{wfsnoise2a}, one should 
note that \inlinecite{1993rtpf.conf..124M} used an approximation for 
an image of a given contrast, using the above notation:
\begin{equation}
\sigma^2_x \approx \frac{5m^2 \sigma^2_{\mathrm {b}}}{n^2_{\mathrm {r}} C 
g\bar{S}_{\mathrm{ADU}}} \frac{(d\theta)^2}{\lambda^2}
\qquad [\mathrm{waves}^2],
\label{wfsn2}
\end{equation}
where $C$ is a factor related to the image contrast and $n^2_{\mathrm {r}}$ is 
the pixel area of each sub-aperture image. The value of $C$ is found by 
dividing Equation (\ref{wfsn}) by Equation (\ref{wfsn2}). Thus

\begin{equation}
C = \frac{\sigma^2_{\mathrm {i}}}{g\bar{S}_{\mathrm{ADU}}},
\label{C}
\end{equation}
for an exposure time of 900$\mathrm{\upmu s}$.
The results were then plotted by scaling $\sigma_{\mathrm {i}}$ and 
$\sigma_{\mathrm{Photon}}$ with exposure time.

\subsection{WFS Error from Telemetry}\label{sec:telem}

In order to verify the above error measurements, the error in the 
SHWFS was found by direct measurement from SHWFS telemetry.
 This was done by taking the power spectrum 
of the total pixel shifts as measured, that is, $\mathrm{shift} = 
\sqrt{x_{\mathrm {shift}}^2 + y_{\mathrm {shift}}^2}$. 
The power spectrum should be taken for data 
with a zero mean, as stated by \inlinecite{marino}. 
So the average 
shift was subtracted from each member of the shift array, for each data set. 

\begin{figure}[t]
\centering
\includegraphics[width=0.55\textwidth,clip,angle=270,origin=bl]{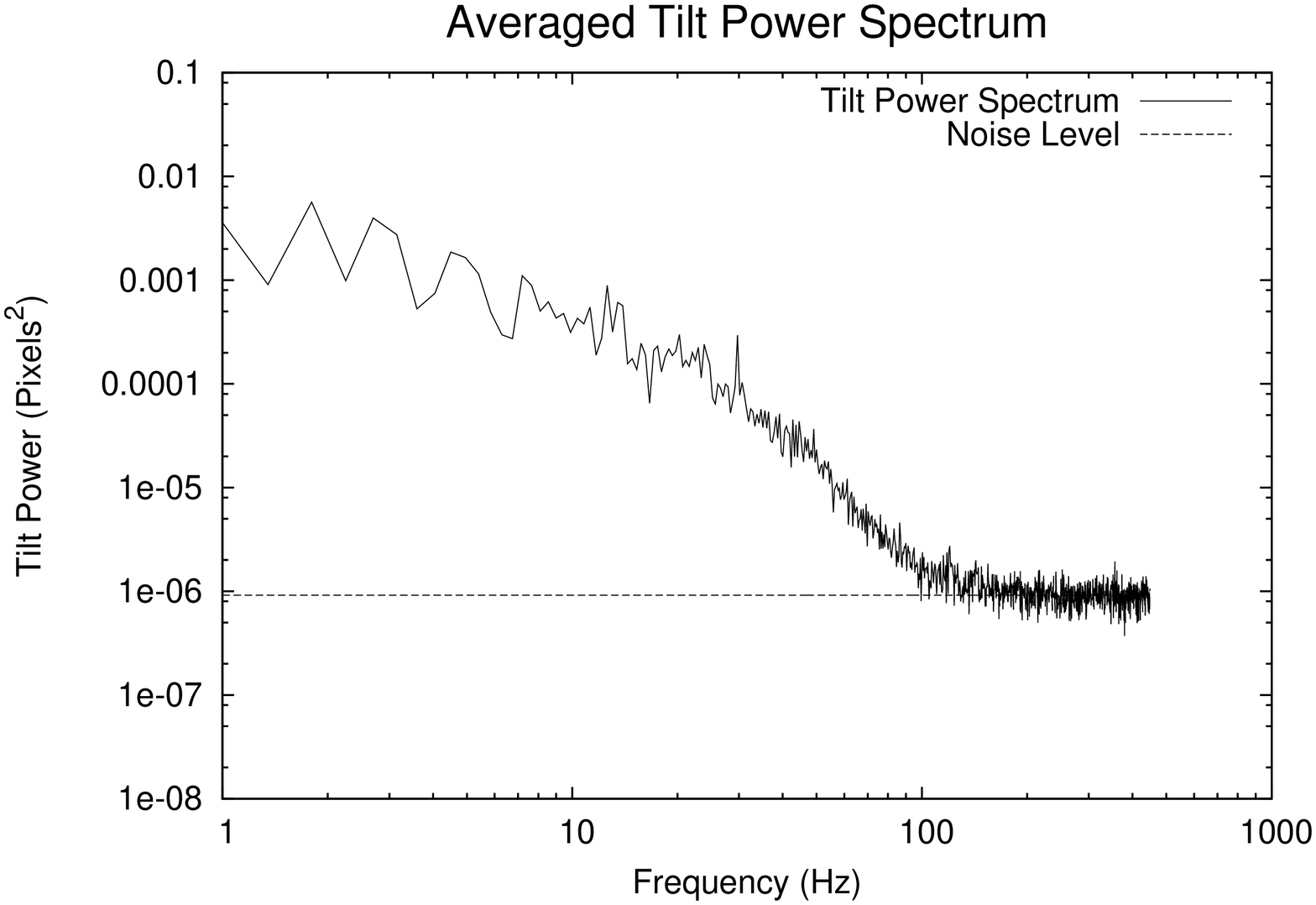}
\includegraphics[width=0.55\textwidth,clip,angle=270,origin=bl]{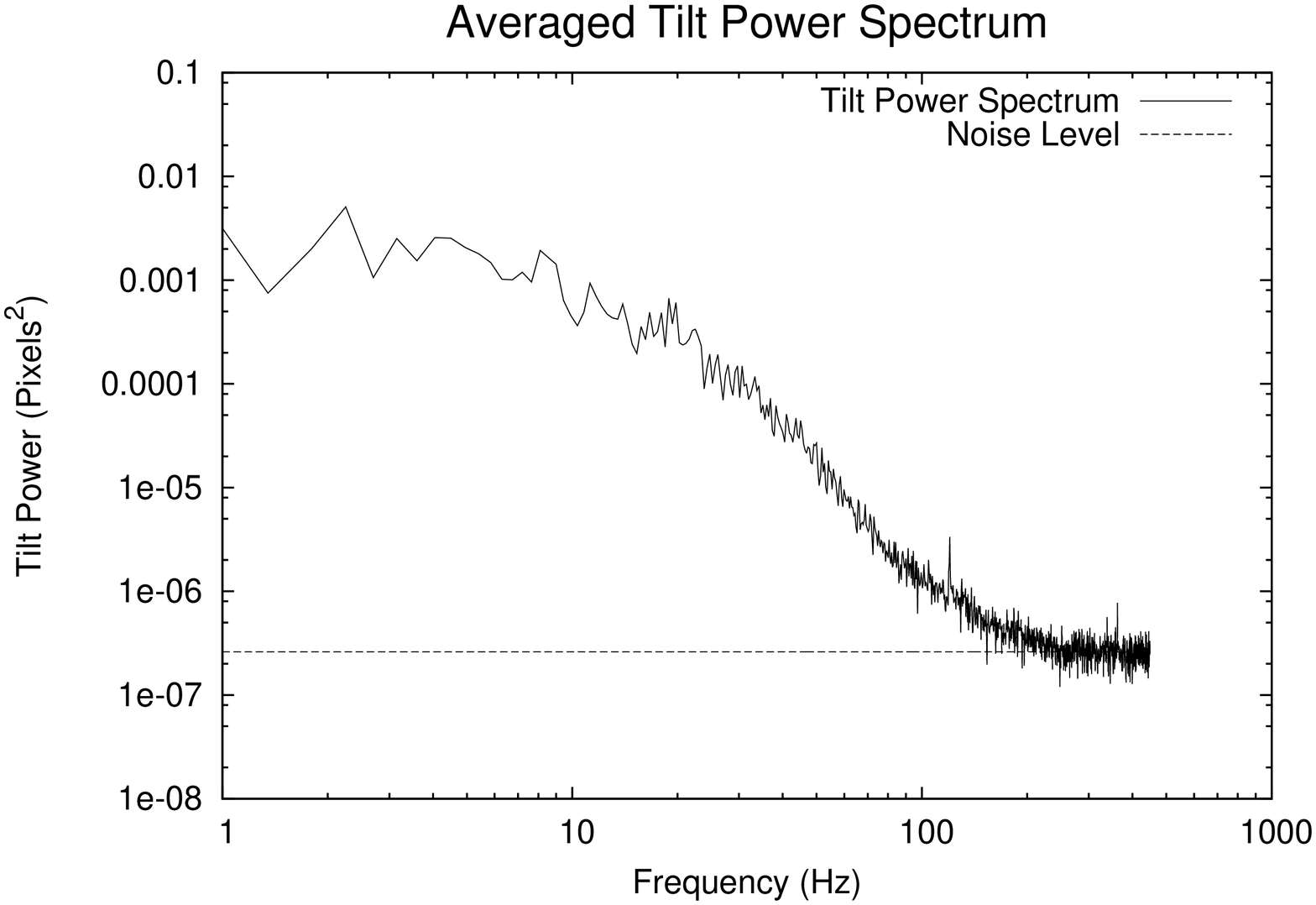}
\caption[]{Noise power spectrum for data taken at 900 Hz, for moderate 
(top panel) and high (bottom panel) contrast prominences.}
\label{pspec1}
\end{figure}

Using the method of \inlinecite{marino}, assuming white noise, we obtain

\begin{equation}
\sigma_{\mathrm{WFS}}^2 = \sum_1^N{\mathrm{Noise\;level}.}
\label{pownoise}
\end{equation}

\noindent Hhere the 'Noise level' is the point where the power spectrum 
becomes flat and $N$ is the total number of exposures in each time series
(see Figure \ref{pspec1}). This becomes

\begin{equation}
\sigma_{\mathrm{WFS}}^2 = \mathrm{Noise\;Level} \cdot N,
\label{pownoise2}
\end{equation}

\noindent which yields the noise variance in terms of $\mathrm {pixels}^2$. 
To find the RMS noise 
in terms of nanometers, the following is applied,

\begin{equation}
\sigma_{\mathrm{WFS\;nm}} = \sigma_{\mathrm{WFS}} \theta  d/(3600 \cdot 57.3). 
\label{pownoise3}
\end{equation} 

\noindent Here $\theta$ is the pixel scale, $0.82"$ per pixel, and 
$d$ is the sub-aperture diameter, 13.07 cm.

\begin{figure}[t]
\centering
\includegraphics[width=0.8\textwidth,clip]{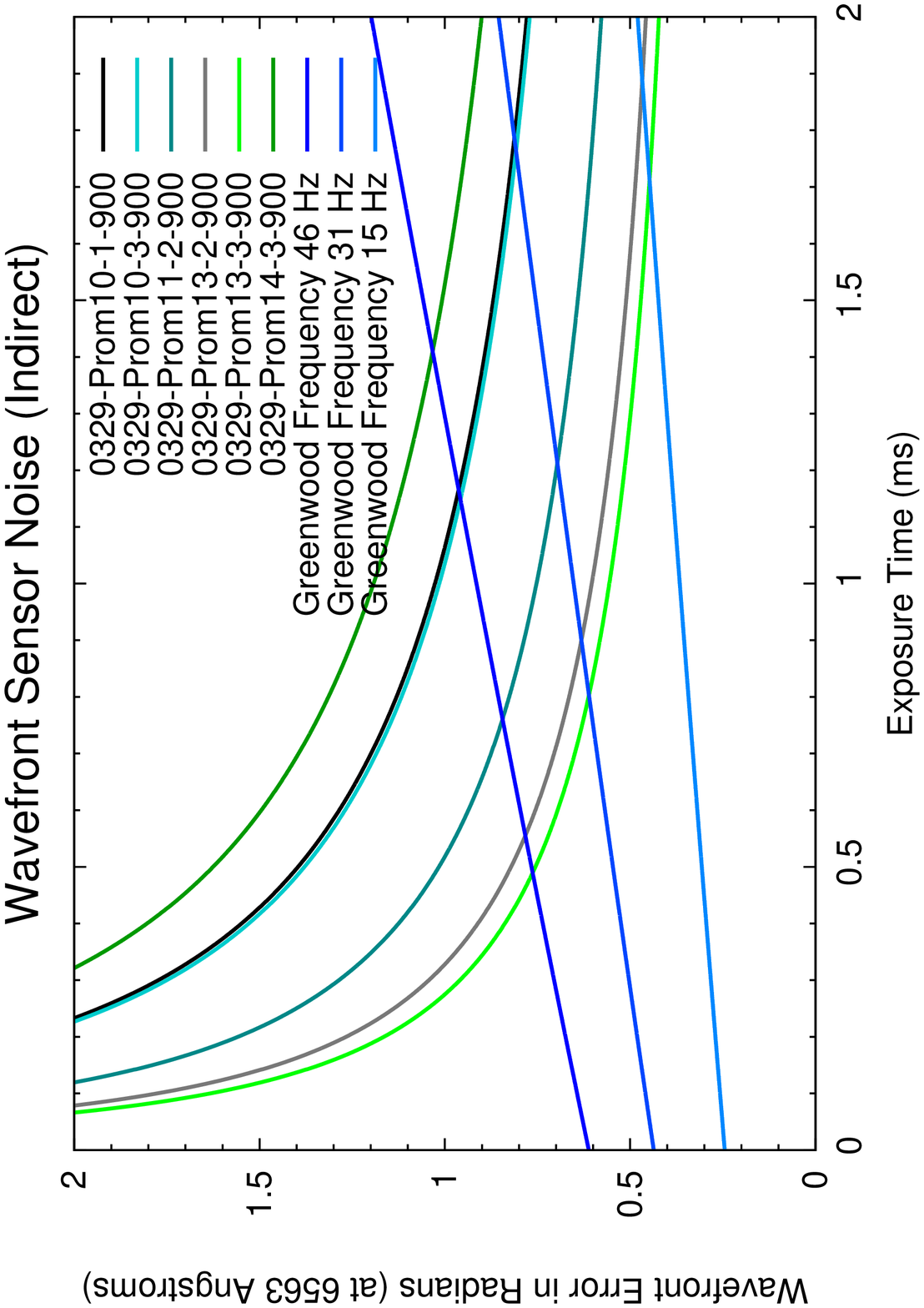}
\includegraphics[width=0.8\textwidth,clip]{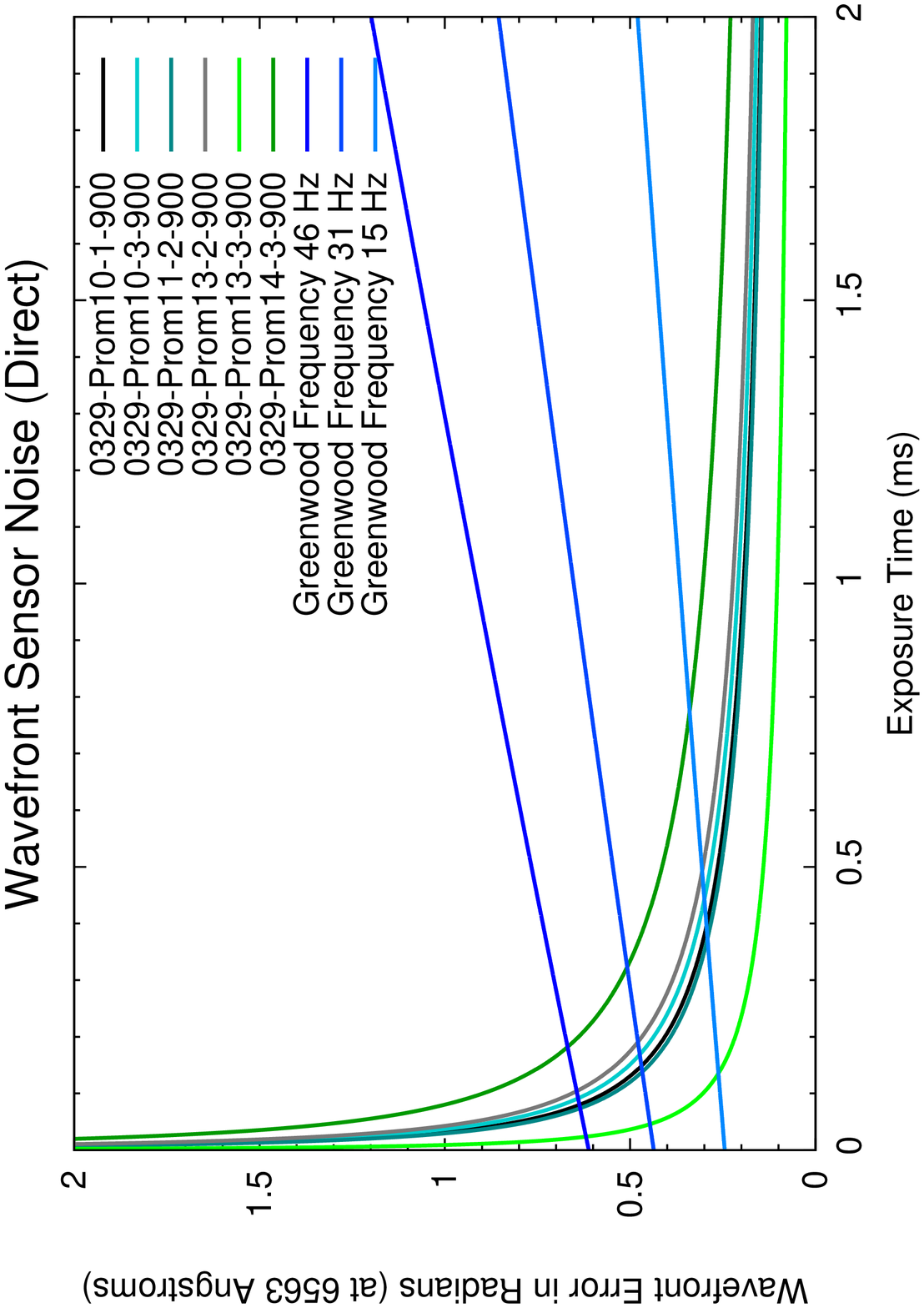}
\caption[WFS and Temporal Noise, Falcon VGA300 HG]
{Top: The solid red to green lines represent the WFS noise for the various 
prominences which were observed.  
The time delay errors (blue), are for 10cm seeing 
(at 5000 {\Ang}, 13.8 cm at 6563 {\Ang}, 
for which these errors are measured), with wind 
speeds of 5, 10, and 15 m s$^{-1}$, from bottom to top, respectively. These 
correspond to Greenwood frequencies of 15Hz, 31Hz, and 46Hz, again at 
{\Ha}. See Section \ref{sec:sr} for an explanation of time-delay error.
Right: Noise calculated from telemetry data.}
\label{wfsnoise2a}
\end{figure}

To find the error in radians, we use

\begin{equation}
\sigma_{\mathrm{WFS\;rad}} = 2\pi \sigma_{\mathrm{WFS\;nm}}/\lambda, 
\label{pownoise4}
\end{equation}

\noindent with $\lambda$ being 6563 {\Ang}.
The ratio between $\sigma_{\mathrm {WFS\;rad}}$ and $\sigma_x$, 
as found in 
Section \ref{sec:ind}, was calculated for each exposure time where 
$\sigma_{\mathrm {WFS\;rad}}$ was 
defined. The value of $\sigma_x$ was then scaled by the average of these 
ratios, for each prominence. 
Equation (\ref{wfsn2}) was then fit to the scaled $\sigma_x$, 
as shown in Figure \ref{wfsnoise2a}. (The error due to 
the rate at which a SHWFS can measure the wavefront is also shown, as 
explained in Section \ref{sec:sr}.) Comparing the result from 
Equation (\ref{wfsn}) with the telemetry measurement, Equation (\ref{wfsn}) is 
found to overestimate the noise by about a factor of 4.

Figure \ref{wfsnoise2a} shows the importance of balancing the speed of 
the wavefront measurements; the shorter the exposures, and thus the 
faster the wavefront measurements, the lower the noise from the 
atmosphere, but the higher the noise from the SHWFS itself. 
The way in which this impacts the quality of the final image in will be shown 
in Section \ref{sec:sr}.

\subsection{Strehl Ratio}\label{sec:sr}

The quality of an image is often measured in terms of its Strehl ratio. 
The Strehl ratio is the ratio of the intensity of the peak of a point 
source image measured by an imperfect optical system against 
the intensity of that 
same object as measured by a perfect optical system. A perfect optical 
system will have a Strehl ratio of 1. Any imperfection in the optics or 
caused by atmospheric distortion will lower the Strehl ratio \cite{har_98,B&W}.
A Strehl ratio above 0.8 is considered diffraction-limited \cite{B&W}. However, 
an image still contains all of the information of a diffraction-limited one, 
only at a lower signal-to-noise ratio, until the Strehl ratio drops 
below about 0.1 \cite{har_98,1998ApOpt..37.4561L}. However, the higher 
the achieved Strehl ratio, the higher the signal-to-noise ratio of the 
data.

\begin{figure}[h]
\centering
\includegraphics[width=0.8\textwidth,clip]{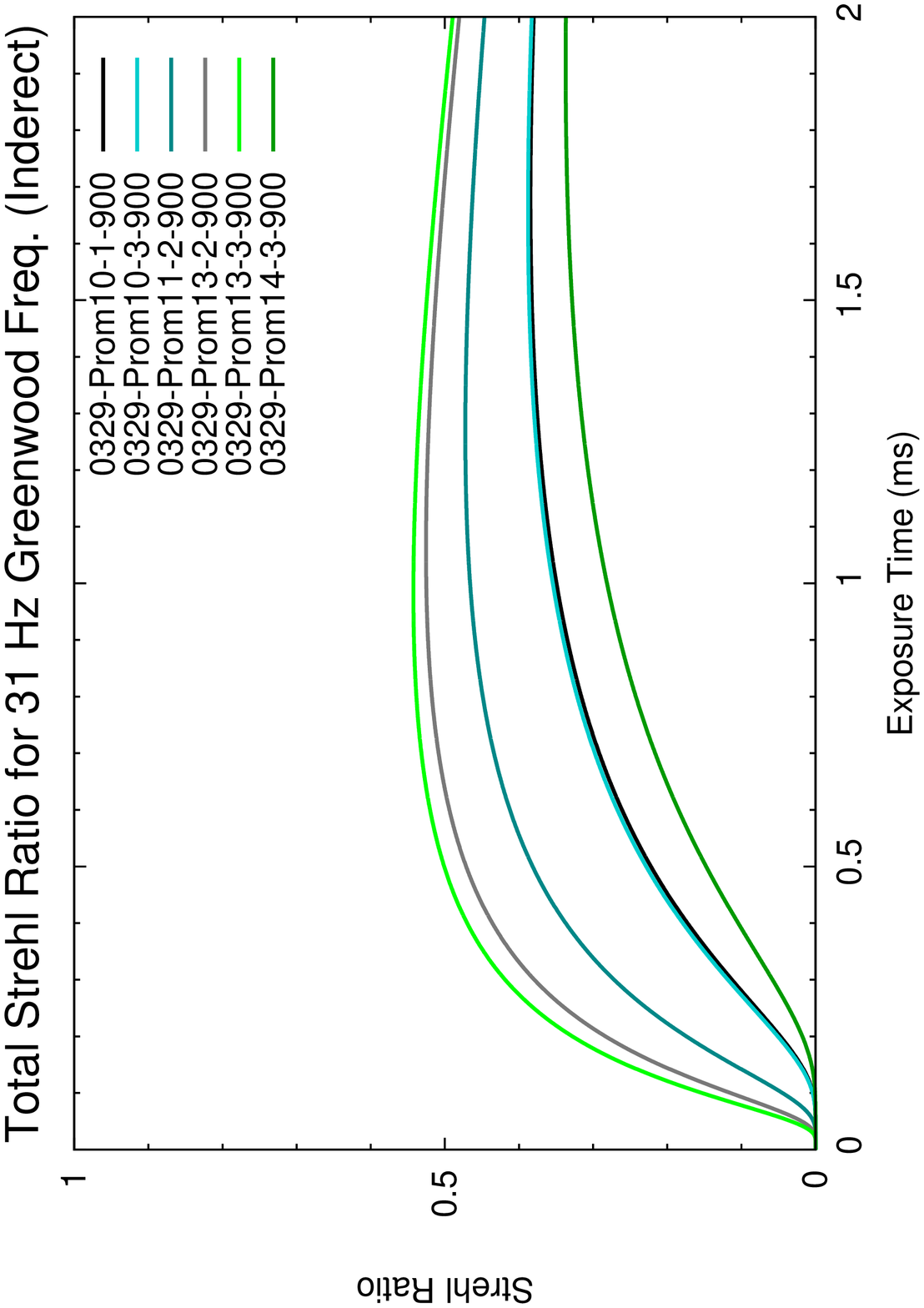}
\includegraphics[width=0.8\textwidth,clip]{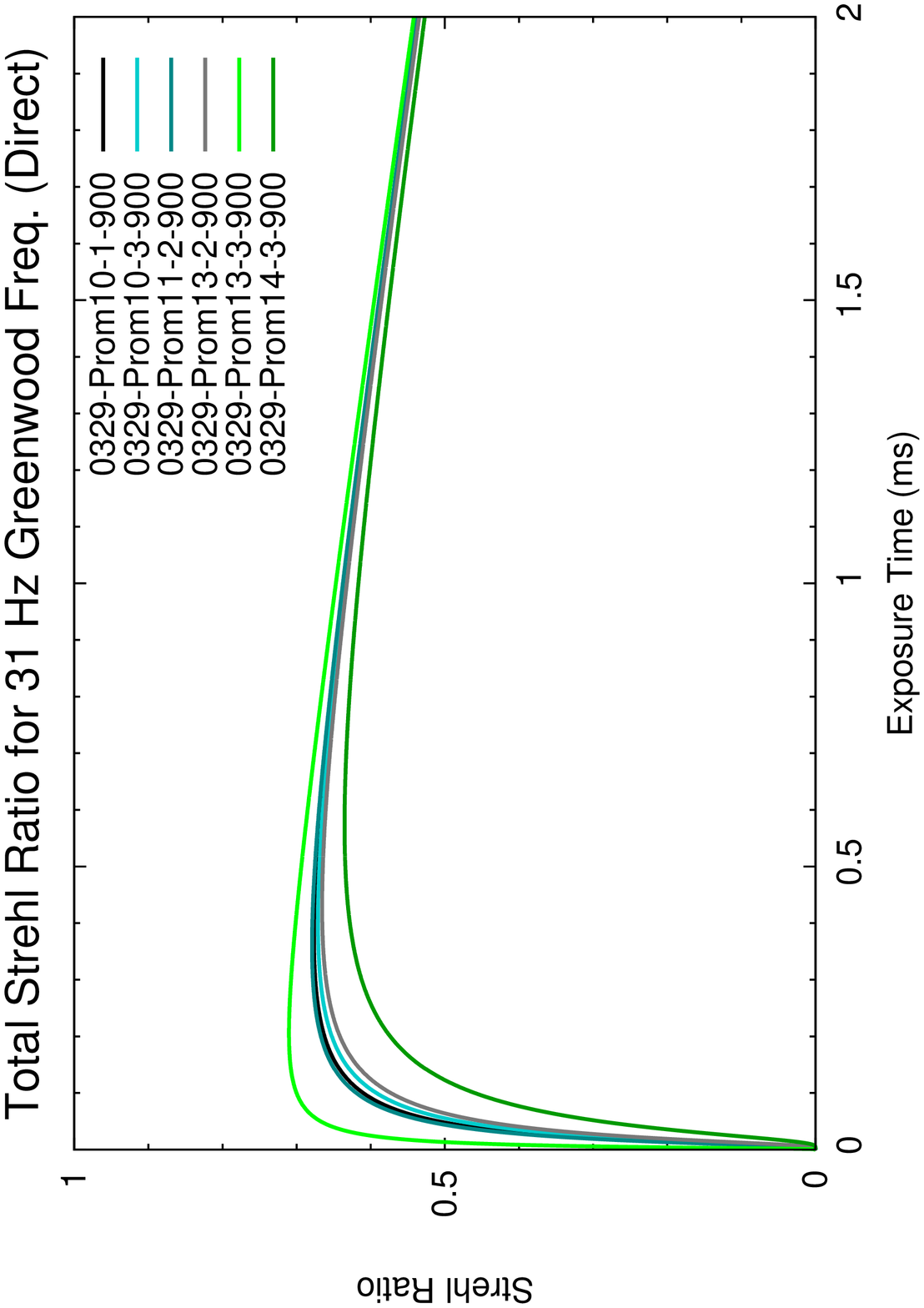}
\caption[Falcon VGA300 Strehl ratio]{Left: Calculated Strehl ratio, taking into 
account all error sources. Only the curves for a Greenwood frequency 
of 31Hz are shown. Right: Calculated Strehl ratio from telemetry data. 
These data are for 8542{\Ang}.}
\label{strehl}
\end{figure}

In order to determine the Strehl ratio that one can expect to achieve from 
an AO system using this SHWFS, it is necessary to 
account for the other primary sources of error: time delay, fitting and 
aliasing errors. For all of the following error estimates, 
a Fried parameter \cite{1965JOSA...55.1427F}, $r_0$, of 10cm, 
at 5000 {\Ang} was assumed (See \inlinecite{har_98} for a more detailed 
discussion of AO error sources.).

To determine the error due to time delay, the functional form 
quoted by \inlinecite{har_98} was used:
\begin{equation}
\sigma^2_{\mathrm{TD}}=28.4(\tau_{\mathrm{s}}f_{\mathrm{g}})^{5/3},
\label{td}
\end{equation}
where $\tau_{\mathrm{s}}$ is the time delay, and $f_{\mathrm{g}}$ is the Greenwood 
frequency ($f_{\mathrm{g}} \approx 0.427 \frac{v}{r_0}$ for a single layer 
atmosphere model, 
which was assumed for this study), $v$ being the wind speed \cite{har_98}.
This error only takes the speed of the camera into account. 

To find $\tau_{\mathrm{s}}$, the delay was assumed to be equal to the exposure 
time plus a constant, pessimistic 500$\upmu$s delay, for wavefront 
reconstruction plus the camera readout, {\it etc}. 
This yields $\tau_{\mathrm{s}} = 500 \mathrm{\upmu s + exposure \; time}$. 
This is similar to the 
method used by \inlinecite{har_98}. Three curves were calculated for $r_0$ of 
10cm at 5000 {\Ang}, which is 13.8cm at 6563 {\Ang}, using the formula 
$r_0' = r_0(\lambda/5000)^{6/5}$ \cite{har_98}. Wind speeds of 
5, 10, and 15m s$^{-1}$ were chosen. Thus, the Greenwood frequencies 
were 15Hz, 31Hz, and 46Hz (see Figure \ref{wfsnoise2a}). The formula 
for $\tau_{\mathrm{s}}$ above does not
take into account the driving frequency of the deformable mirror (DM), which 
was unknown when this analysis was performed. 

The forms of fitting and aliasing errors were used, as described by 
\inlinecite{2011LRSP....8....2R}, where the fitting error is
\begin{equation}
\sigma^2_{\mathrm {F}} = 0.28\left(\frac{d}{r_0}\right)^{5/3}.
\label{fit}
\end{equation}
The aliasing error is
\begin{equation}
\sigma^2_{\mathrm {A}} = 0.08\left(\frac{d}{r_0}\right)^{5/3},
\label{alias}
\end{equation}
where $d$ is the sub-aperture size.

In Figure \ref{strehl}, the resultant 
Strehl ratios, given the above errors, are shown. 
The approximation $\mathrm{Strehl \; ratio} = {\mathrm e}^{-\sum_i 
\sigma_i^2}$ was used
 to find these Strehl ratios \cite{har_98}. All of the Strehl ratios were 
calculated at 8542 {\Ang}. To convert the SHWFS and time delay errors 
at 6563 {\Ang}, which were measured in radians, the authors took note that 
$1\;\mathrm{wave} = 2\pi \;\mathrm{radians}$ at a given wavelength, so
\begin{equation}
\sigma_{8542\;\mathrm{in\;radians}} = \frac{6563}{8542}\sigma_{6563\;\mathrm{in\;radians}.}
\label{convert}
\end{equation}
Figure \ref{strehl} shows that one can expect to see Strehl ratios of between 
0.6 and 0.7, at 8542 {\Ang}, given a Fried parameter of 10cm and 10m s$^{-1}$ 
wind.

There are a few important facts that can be gleaned from Figure \ref{strehl}. 
(1) The faster the SHWFS is able to measure the wavefront, the better the 
achieved Strehl ratio. (2) Using faint, low contrast prominences for sensing 
the wavefront lowers the maximum achievable Strehl ratio. (3) For a fainter 
prominence, the maximum achievable Strehl ratio occurs at a slightly slower 
frame-rate than for a brighter prominence. (4) This SHWFS is limited by 
hardware to a speed of around 900Hz, which corresponds to 
an exposure time of about $900\upmu {\mathrm s}$. Given a faster camera, 
slightly better Strehl ratios could be achieved, but the achievable Strehl 
ratio decreases slowly, to the right of its maximum point. 

\section{Experimental Setup and Procedure}\label{sec:setup}

The off-limb AO SHWFS, as tested in Section \ref{sec:pred}, 
was integrated into AO bench of the DST. 
A 97-actuator, 
Xinetics deformable mirror (DM) and a 
tilt-tip mirror (TTM) were utilized. 
Wavefront reconstruction is accomplished by utilizing a customized 
implementation of the 
Kiepenheuer-Institute Adaptive Optics System (KAOS), which was originally 
coded for the German vacuum tower telescope (VTT) and was rewritten 
to operate the 
GREGOR telescope \cite{2012AN....333..863B}. It was customized to utilize 
the layout of the DM as well as the layout, exposure time, {\it etc.}, of the 
SHWFS.
This implementation of KAOS is run on an off-the-shelf computer, 
utilizing an Intel i7, quad-core 
processor. This is interfaced to the DM using a Xinetics, 
Gen III chassis, driving 144 
channels. Of the 144 channels, 97 channels are used for the DM, 2 for the TTM, 
and the remaining channels are unused.

\begin{figure}[h]
\centering
\includegraphics[width=0.75\textwidth,clip]{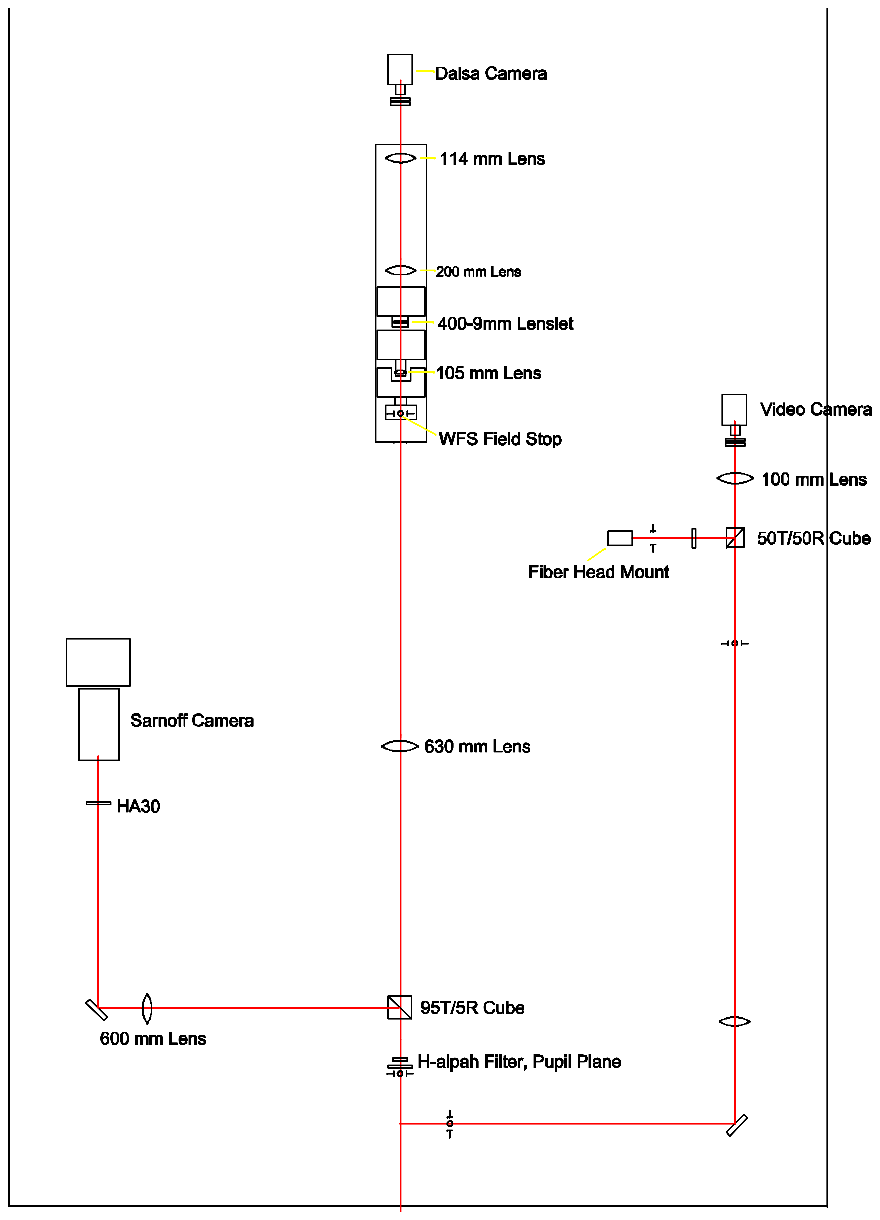}
\caption{The 
experimental setup on Port 2 of the DST. The WFS is at top, with 
the imaging camera at left. An interferometer can be used to verify the 
flatness of the DM. It is to the right of the WFS and is used by 
diverting all light from the WFS, via a mirror, along the rightmost 
red path.}
\label{fig:port2-closeup}
\end{figure}

Tests were performed on Port 2 of the DST.  
The setup consists 
of a beam-splitter, which diverts 95\% of the incoming {\Ha} light to the 
SHWFS, leaving the remainder to feed the imaging camera. The rest is 
fed to the SHWFS. The bench optics shown are required to create a $5 \times 5$ 
SHWFS array with sub-aperture FOV of $30''$ and pixel scale of 
$0.82''$ per pixel
(see Figure \ref{fig:port2-closeup}). For these tests, the camera was 
run at a frame rate of $~860$Hz. This is slightly slower than the 
model frame rate, but it allowed for a border of a few pixels around the 
sub-aperture images ($174 \times 174$ pixels are now being read-out). 
This allows 
the KAOS system to compensate for slight misalignments in the optical system. 
Since the system is running at a slightly slower frame rate than used in 
Section \ref{sec:pred}, an exposure time of 1ms could be used. This lowered 
the noise coming from the SHWFS, at the expense of slightly more 
time-delay error.

\subsection{Procedure}\label{sec:proc}

The KAOS system is capable of recording data which contains raw measurements 
from the SHWFS as well as the calculated errors in the reconstructed 
wavefront. KAOS records the commands sent to the TTM and the DM. It 
calculates the total atmospheric error in each mode, given the residual 
error and the commands sent to the DM.  
These data and their caveats are explained below.

When KAOS reconstructs the wavefront, it does so by utilizing Karhunen-Lo\`eve 
(K-L) modes \cite{1976JOSA...66..207N, 1995JOSAA..12.2182D}. The degree of 
correction that can be 
achieved depends upon the number of modes that can be utilized, which 
in turn depends upon the resolution to which the SHWFS can sample the 
wavefront \cite{1976JOSA...66..207N, har_98}.
Even if those K-L modes can be perfectly determined, there is a 
residual error in the wavefront, corresponding to the infinite number of 
K-L modes which were not sensed \cite{1976JOSA...66..207N,1995JOSAA..12.2182D}.
K-L modes are ranked in various orders. Each mode in each consecutive 
order contributes less to the total error in the wavefront. Thus, the 
correction of a few modes can correct most of the distortion in the incoming 
wavefront \cite{1976JOSA...66..207N,1995JOSAA..12.2182D}. 
This setup can sense 20 K-L modes. \inlinecite{1976JOSA...66..207N} has tabulated 
the residual error expected after correcting 20 Zernike modes, which are 
similar to K-L modes, but with slightly higher residuals errors
\cite{1976JOSA...66..207N,1995JOSAA..12.2182D}. 

The other major caveat is 
the way in which KAOS arrives at its measurements of the total wavefront 
error in each K-L mode. It is done by noting the residual error in each 
mode and determining how much correction is being applied in each mode 
via the TTM and the DM and combining the two. If the pixel scale of the 
SHWFS was measured incorrectly, both the residual and total error measurements 
will be wrong. However, these measurements of the pixel 
scale have been ensured to be accurate.
The DM used was a spare at the DST, and its 
neutral state was not exactly flat. This coupled with any static errors 
in the system, which would be corrected by the DM, affects the total 
wavefront error measurements.

To compare the measured performance of the off-limb solar AO system with the 
expected performance, shown in Figure \ref{strehl}, there are a few parameters 
which are of interest: $r_0$, $B_{\mathrm {cl}}$ (the closed-loop bandwidth), 
and the Strehl ratio of 
the system. If the AO loop is treated like an RC filter, 
where higher update frequencies are attenuated, then $B_{\mathrm {cl}}$ can be 
defined as the $0{\mathrm{db}}$ bandwidth. This is the highest frequency at which 
accurate wavefront corrections can be made \cite{roddier}. 
See the explanation of Strehl ratio in Section \ref{sec:sr}.

The value of $r_0'$ can be calculated from the formula given by 
\inlinecite{1976JOSA...66..207N},

\begin{equation}
\bar{\phi}^2 = C \cdot (D/r_0')^{5/3}.
\label{residual}
\end{equation}

\noindent Here $D$ is the diameter of the telescope and $\bar{\phi}^2$ is 
the variance wavefront error, in radians, averaged with time. 
$C$ is a constant which is defined by the expected relative variance in a 
given mode, for a given turbulence model 
\cite{1976JOSA...66..207N,1995JOSAA..12.2182D}.
The average variance of the first three K-L modes, after the two modes 
corresponding to tip-tilt, 
were used, since these have the same approximate variance of 
$0.023927 (D/r_0')^{5/3}$ \cite{1995JOSAA..12.2182D}. Also, tip-tilt 
mode measurements are sensitive to vibrations in the room, {\it etc}. 
The quantity 
$r_0'$ was calculated by 
finding the average wavefront error over a series of 10,000 frames, 
or approximately 12 s. With $r_0'$ known, it is trivial to find 
$r_0$. The Strehl ratio can be approximated by\cite{har_98} 
\begin{equation}
{\mathrm e}^{-1\times[(\sum_{i} \bar{\phi}_{i}^2) + {\mathrm {residual}}]}.
\label{eq:sr}
\end{equation}

\noindent Here $\sum_i \bar{\phi}_i^2$ is the sum of the residual errors 
for each mode over the time series, 
calculated by KAOS, and the residual is approximated here as \\
$0.0208 (D/r_0')^{5/3}$ \cite{1976JOSA...66..207N}. This is the residual for 
Zernike polynomials, but it is only slightly more than that for K-L modes
\cite{1976JOSA...66..207N,1995JOSAA..12.2182D}.

\begin{figure}[t]
\centering
\includegraphics[width=0.55\textwidth,clip,angle=270,origin=bl]{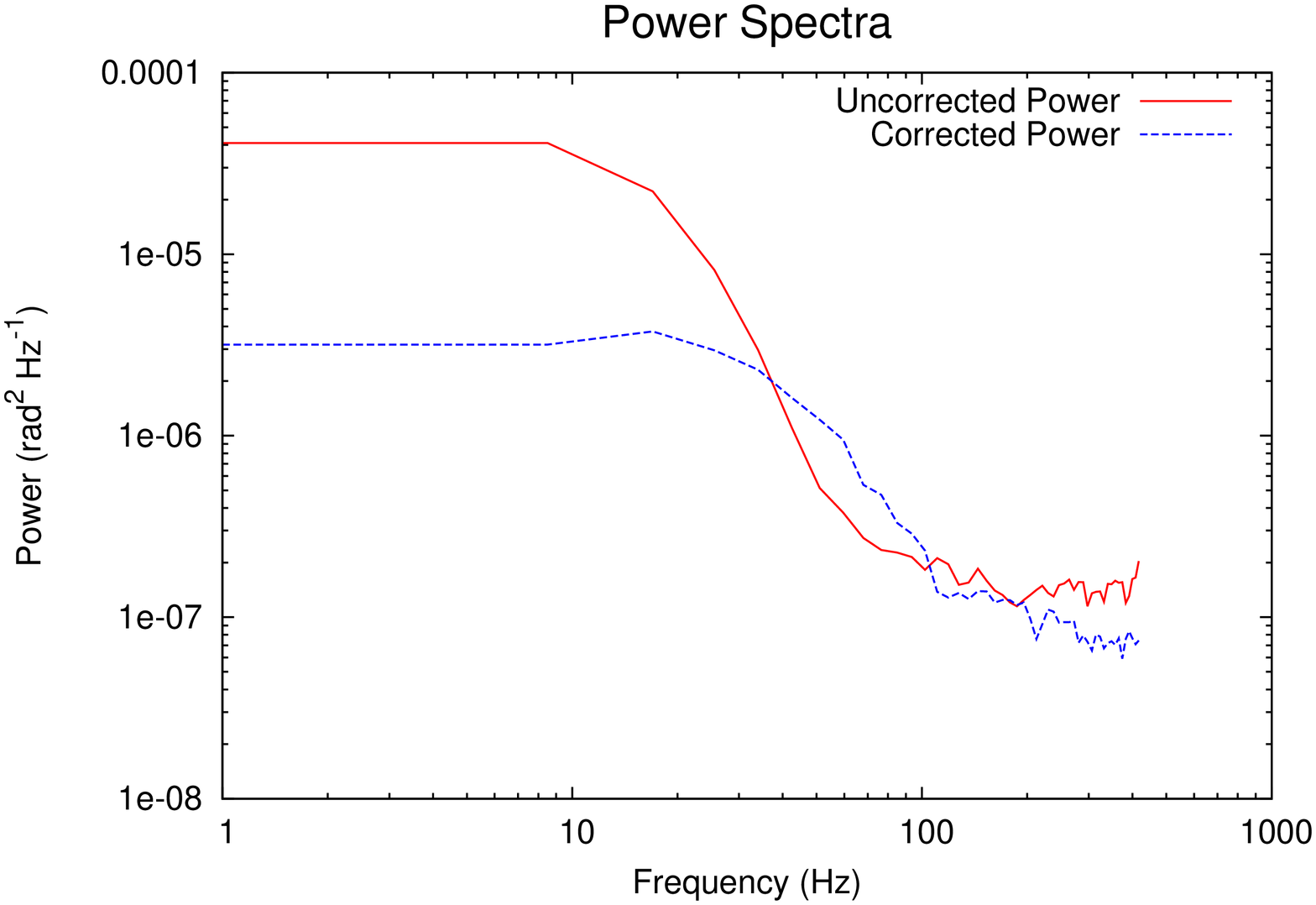}
\caption{The power spectra from which $B_{\mathrm {cl}}$ was calculated. The 
approximate $0{\mathrm{db}}$ bandwidth value was found by interpolation.}
\label{fig:whatever}
\end{figure}

Finding $B_{\mathrm {cl}}$ is a bit more involved. It is found by taking the 
power spectrum of the residual error for a given mode with the AO system 
off, and plot it against the power spectrum of the same mode with the 
AO system on (see Figure \ref{fig:whatever}). 
The point at which they first cross is the $0{\mathrm{db}}$ $B_{\mathrm {cl}}$.
 The Welch method was used to plot the 
power spectra \cite{1967ITAE...15...70W}.

\section{Results}\label{sec:res}

The telemetry data taken according to the procedure found in 
Section \ref{sec:proc} were analyzed.   These data were taken with the AO 
system correcting for 20 K-L modes. The $0{\mathrm{db}}$ bandwidth was found to 
be approximately $38.8$Hz (see Figure \ref{fig:whatever}). This value depends 
upon the degree of smoothing of the power spectra, so it is only 
a ''ballpark figure''.

A sample of the $r_0$ values and Strehl ratios, which were calculated using 
the above method, is shown in Table \ref{tbl:results}.
The data represent a temporal average 
over a period of 12 s. One can note the large standard 
deviation among the $r_0$ values. This is due to the rapidly changing 
atmosphere. Since total Strehl ratio depends 
upon the Greenwood frequency, which was not measured with 
the available data, it is no surprise that 
there is variation in measured Strehl ratios, given similar $r_0$ values. 
Even given this fact, the below results 
measure favorably against the above predictions, as shown in 
Figure \ref{strehl}. It should be noted that even with the very large values 
of $r_0$ present in some of the data sets, correction of 20 K-L modes 
provides higher Strehl ratios than tip-tilt correction alone. This 
can be seen by using Equation (\ref{residual}),
\begin{equation}
\bar{\phi}^2 = C \cdot (D/r_0')^{5/3}.
\end{equation}
If only tip-tilt errors are corrected, the constant $C$ is equal to $0.134$ 
\cite{1995JOSAA..12.2182D}. 
This yields a residual error at 5000 {\Ang}, at $r_0=17.5{\mathrm {cm}}$, of $1.56$ 
radians$^2$. This,

\begin{table}[h]
\tiny
\begin{tabular}{ccccc}
\hline
$r_0$ cm, 5000 {\Ang} & Standard deviation & Strehl, 8542 {\Ang}& 
Standard deviation & Strehl, 6563 {\Ang} \\
\hline
15.9184 & 7.5325 & 0.7926 & 0.0652 & 0.6773 \\ 
14.9362 & 6.9773 & 0.7884 & 0.0610 & 0.6709 \\ 
17.4868 & 9.2392 & 0.8456 & 0.0395 & 0.7536 \\ 
9.1684 & 4.7710 & 0.6605 & 0.0609 & 0.4978 \\ 
6.4664 & 3.1955 & 0.4789 & 0.0713 & 0.2911 \\ 
14.0340 & 6.7349 & 0.7797 & 0.0791 & 0.6601 \\ 
12.1283 & 6.4545 & 0.7635 & 0.0497 & 0.6346 \\ 
6.3531 & 3.2902 & 0.4312 & 0.0820 & 0.2457 \\ 
12.4115 & 8.7072 & 0.7327 & 0.0707 & 0.5937 \\ 
27.7547 & 16.9233 & 0.9097 & 0.0382 & 0.8527 \\ 
46.9440 & 24.7305 & 0.9000 & 0.0707 & 0.8397 \\ 
7.7688 & 6.6223 & 0.6098 & 0.0754 & 0.4366 \\ 
6.5431 & 3.9170 & 0.4841 & 0.0989 & 0.3001 \\ 
12.5479 & 6.7247 & 0.7827 & 0.0560 & 0.6624 \\ 
13.3796 & 7.0181 & 0.7954 & 0.0510 & 0.6803 \\ 
4.8167 & 2.5083 & 0.4631 & 0.0125 & 0.2716 \\ 
15.0399 & 7.6358 & 0.6131 & 0.1380 & 0.4499 \\ 
6.9543 & 3.8224 & 0.2626 & 0.1305 & 0.1192 \\ 
5.2454 & 2.8631 & 0.1084 & 0.0942 & 0.0332 \\ 
8.1936 & 4.3825 & 0.2846 & 0.1346 & 0.1349 \\ 
7.2867 & 3.7827 & 0.3161 & 0.1220 & 0.1549 \\ 
\hline
\end{tabular}
\caption{A sampling of $r_0$ and Strehl ratios, as 
calculated in Section \ref{sec:proc}. The standard deviations 
quoted show the spread in the data over 12 s: Seeing is 
often quite variable.}
\label{tbl:results}
\end{table}

\begin{figure}[h]
\centering
\includegraphics[width=0.85\textwidth,clip]{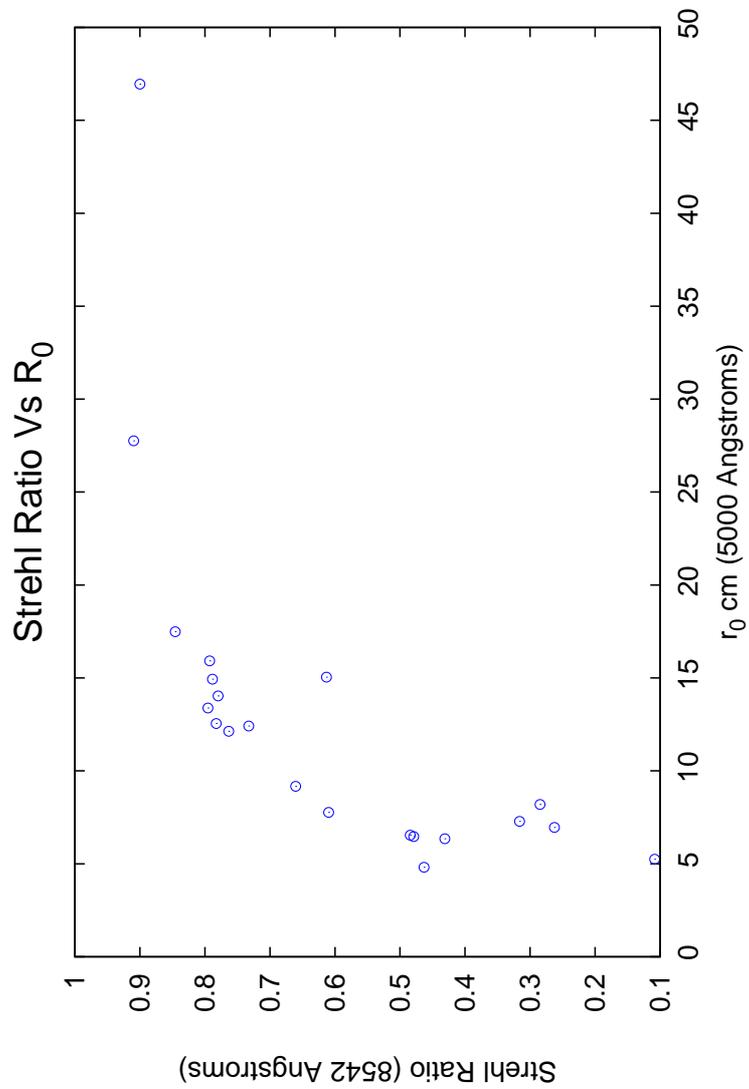}
\caption{A graph of Strehl ratio {\it vs.} $r_0$ value. This takes into account the residual wavefront errors plus the expected variance left-over  from only correcting a finite number of modes (Noll, 1976). Note that for  $r_0 \approx 10$cm, the Strehl ratio is between 0.6 and 0.7, bettering the predictions in Section 2.3}
\label{fig:r0}
\end{figure}
\clearpage
\noindent converting to radians$^2$ at 8542 {\Ang}, as 
in Equation (\ref{convert}), and calculating the Strehl ratio:
\begin{equation}
{\mathrm {Strehl \; ratio}} = {\mathrm{e^{residual \;variance}}},
\end{equation}
yields the maximum possible Strehl ratio at 8542 {\Ang}, only correcting for 
tip-tilt errors of $0.59$. The maximum possible Strehl ratio, 
given correction of 20 modes is 

\noindent $0.92$ using a value of $C$ of $0.0208$ \cite{1976JOSA...66..207N}. 
This system achieved $0.85$. 
So theoretically the improvement is small for excellent seeing, but when 
$r_0$ is $7.7$, the maximum possible 
Strehl ratio is $0.12$, using only tip-tilt corrections. The maximum possible 
Strehl ratio using 
20 K-L modes is $0.72$; this system achieved an estimated Strehl ratio of 
$0.6$.

\begin{figure}[t]
\centering
\includegraphics[width=3.8cm,clip]{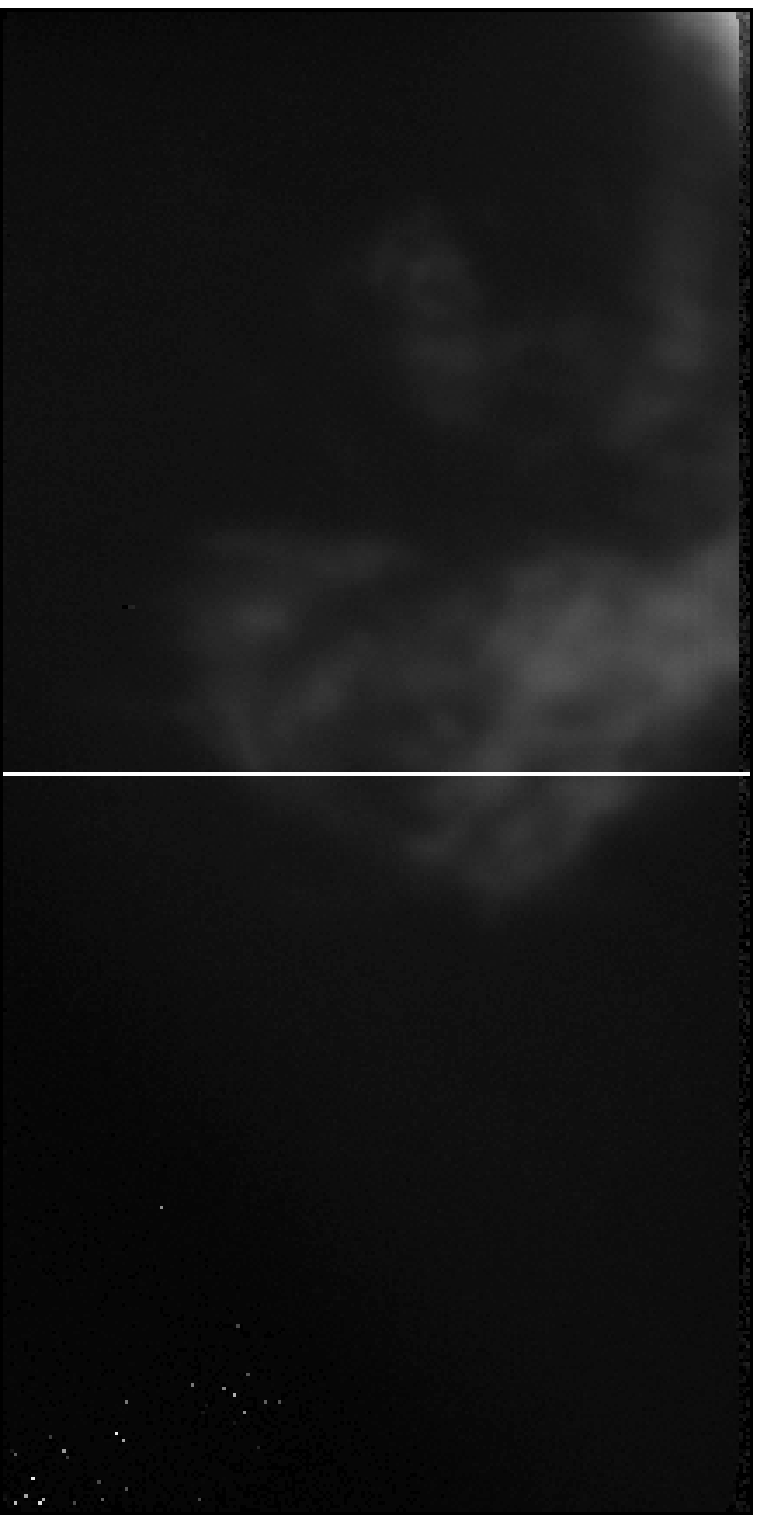}
\includegraphics[width=3.8cm,clip]{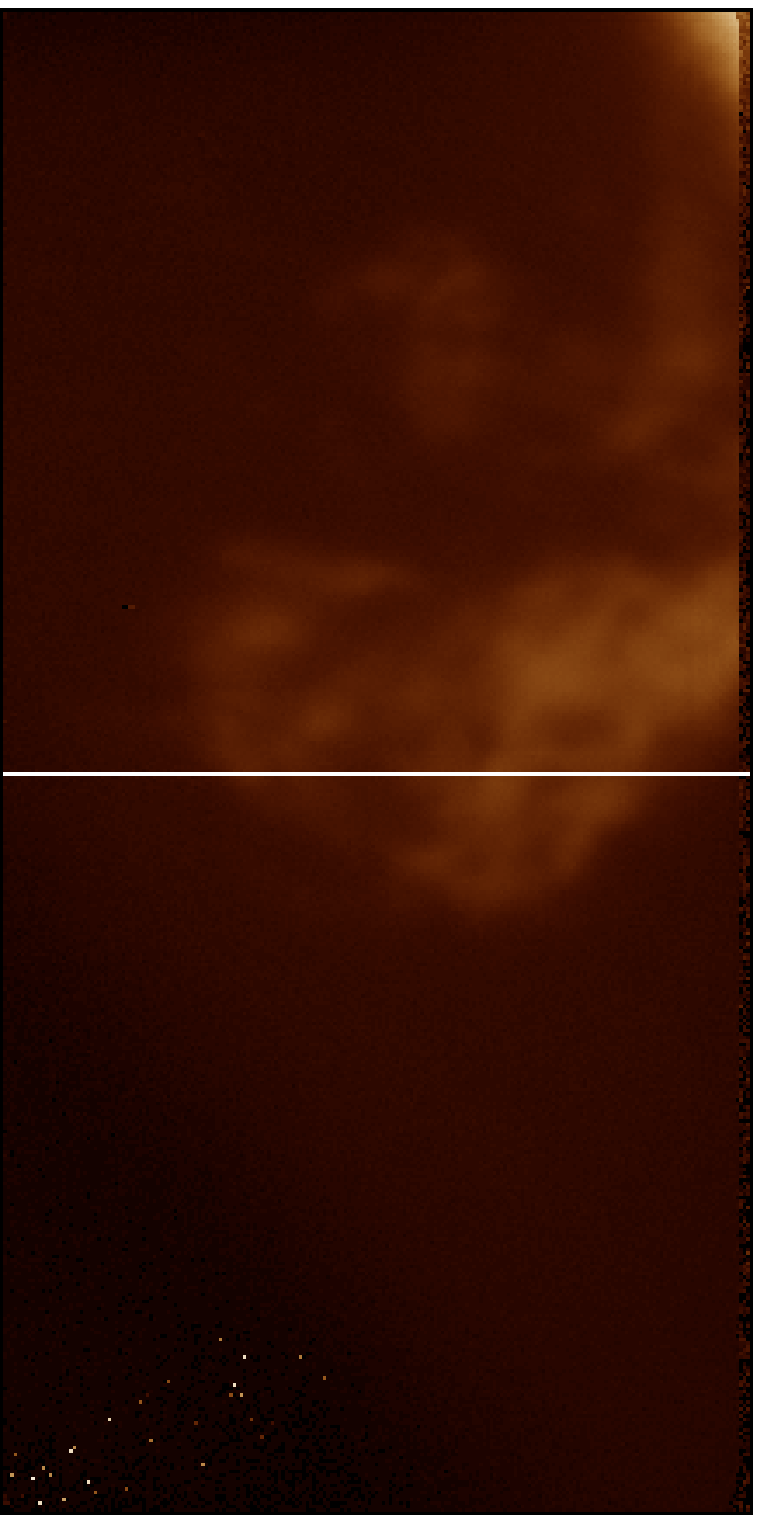}
\includegraphics[width=3.8cm,clip]{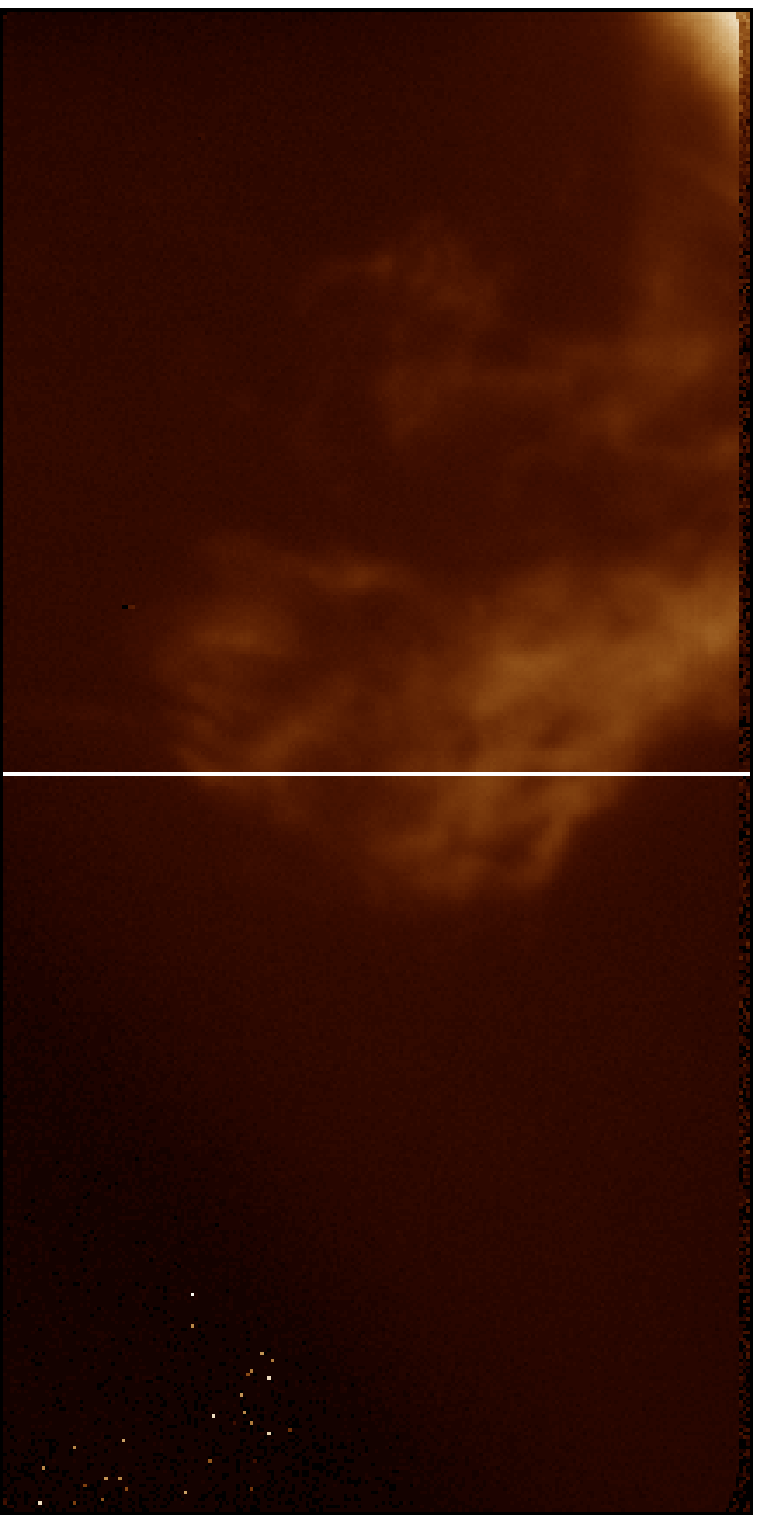}
\caption{The above images were taken on 10 December 2013, 
between 16:50 and 16:55 UTC. Each image is the average of 100 
frames, with a cadence of $4s^{-1}$. Each series was taken at 
6563 {\Ang}. They show what a long exposure 
would look like during moments of moderate seeing, with the AO system 
totally off (left), only correcting for tip-tilt errors (center), and with 
the AO system correcting for 20 modes. $r_0$ varied from between about 15cm 
and 17cm, when these images were taken. (The second line in Table 
\ref{tbl:results} was taken a few moments after the right hand image was 
taken.) Cross sections of each image are shown by the white line. Pixel 
values for each will be shown in Figure \ref{fig:xsctn}}
\label{fig:correction}
\end{figure}
Also note that when the Strehl ratio is highest, its standard 
deviation is lowest, which points to the stability of the 
off-limb solar AO system, 
during times of good seeing. The 5000 {\Ang} $r_0$ and 8542 {\Ang} 
Strehl ratios from Table \ref{tbl:results} are plotted in Figure \ref{fig:r0}. 
Note that for $r_0$ around 10cm, the Strehl ratio is between 0.6 and 0.7, 
slightly exceeding the above predictions. This may be due to the very 
pessimistic time-delay error estimation. The spread in $r_0$ {\it vs.} Strehl 
ratio is almost certainly due to the variation in Greenwood frequency, but 
this is seen to be a secondary effect. Recall that in Section \ref{sec:proc}, 
it was stated that there are error sources that these measurements do 
not take into account. Therefore, the Strehl ratios in Table \ref{tbl:results} 
and Figure \ref{fig:r0} will almost certainly be higher than the 
true Strehl ratios, as might be measured at the final prominence image.
\begin{figure}[t]
\centering
\includegraphics[width=5cm,clip,angle=270,origin=bl]{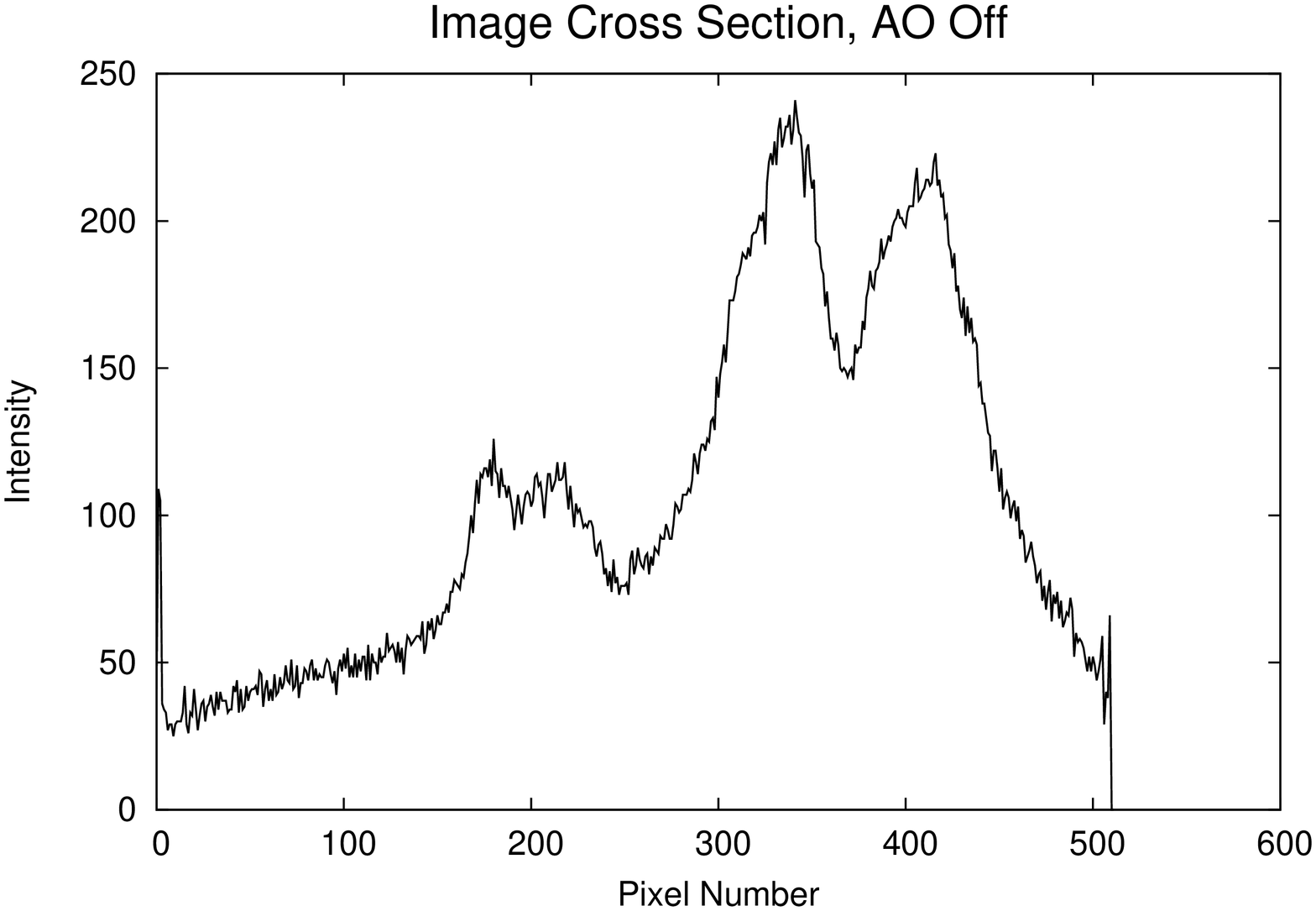}
\includegraphics[width=5cm,clip,angle=270,origin=bl]{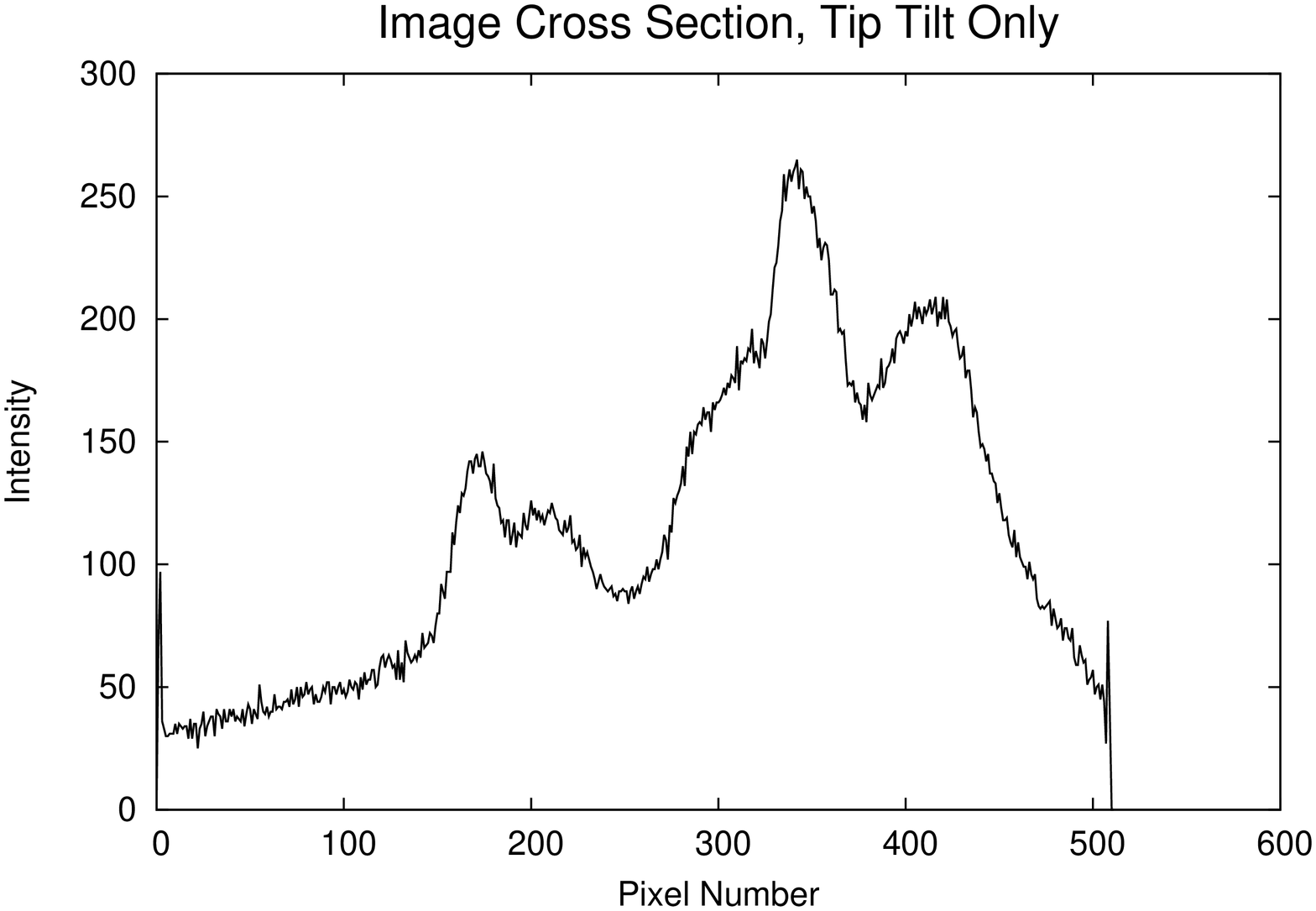}
\includegraphics[width=5cm,clip,angle=270,origin=bl]{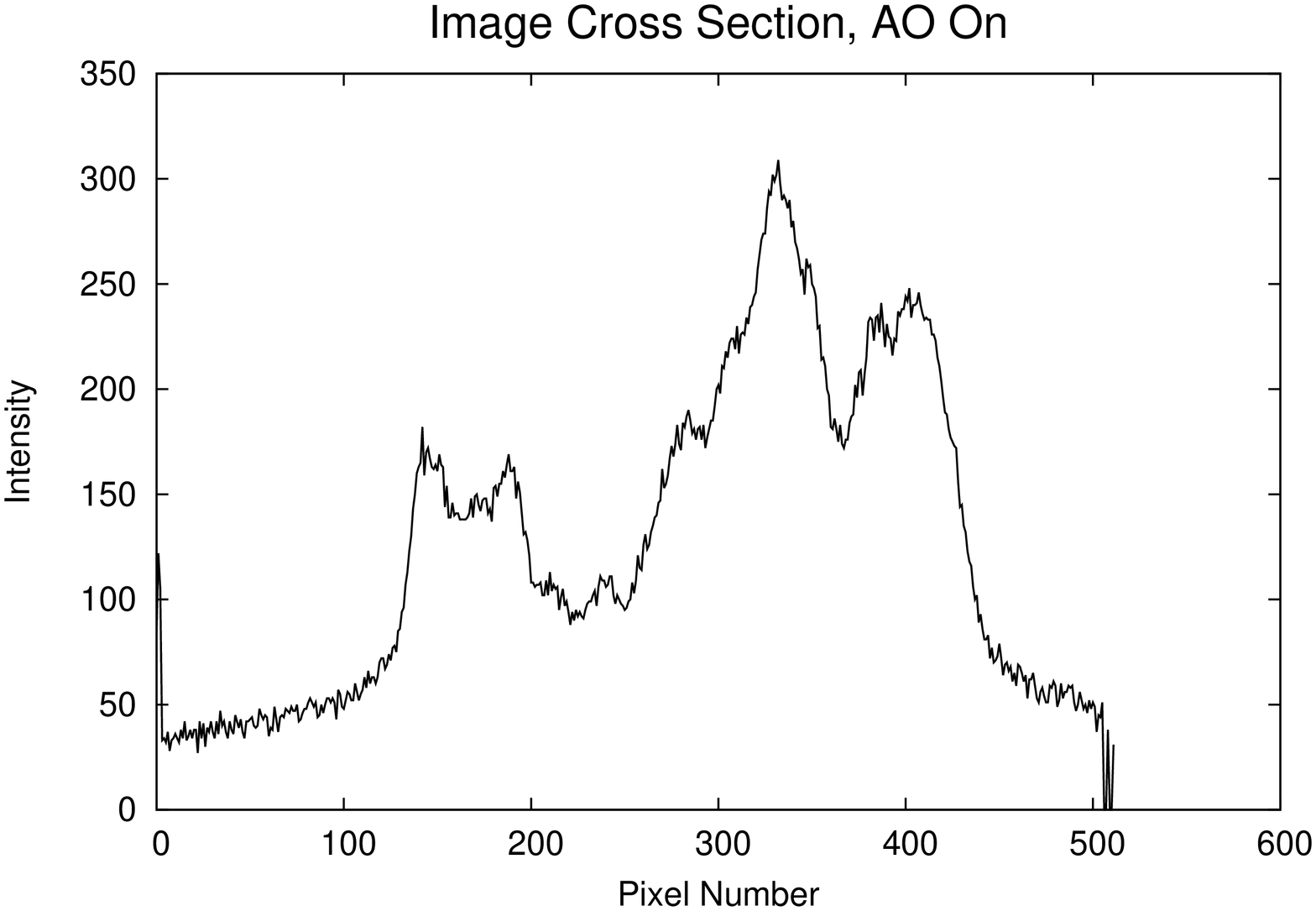}
\caption{Cross sections taken of each image image in Figure 
\ref{fig:correction}. The increase in sharpness for the fully corrected image 
is depicted by the sharp edges shown in the graph. This indicates that much 
finer details are visible.}
\label{fig:xsctn}
\end{figure}
Figure \ref{fig:correction} depicts a few images which show the dramatic 
improvement that the off-limb solar AO system can provide. 
The images are each made from 
a series of 100 frames with an exposure time of 200ms each. They were taken 
at a frame rate of 4s$^{-1}$ and span 25s. To better show 
the resolution improvement made in this figure, their cross sections 
are plotted in Figure \ref{fig:xsctn}. These images and cross sections 
show that even during good seeing, long integrations, which will be required 
for spectro-polarimetry, are vastly clearer with the off-limb solar AO system 
fully operating. During very bad seeing, the tip-tilt mirror can be used 
alone, which is still an improvement over limb tracking, the standard method 
for tracking prominences on the DST (Kevin Reardon, private communication).

\begin{figure}[t]
\centering
\includegraphics[width=0.8\textwidth,clip]{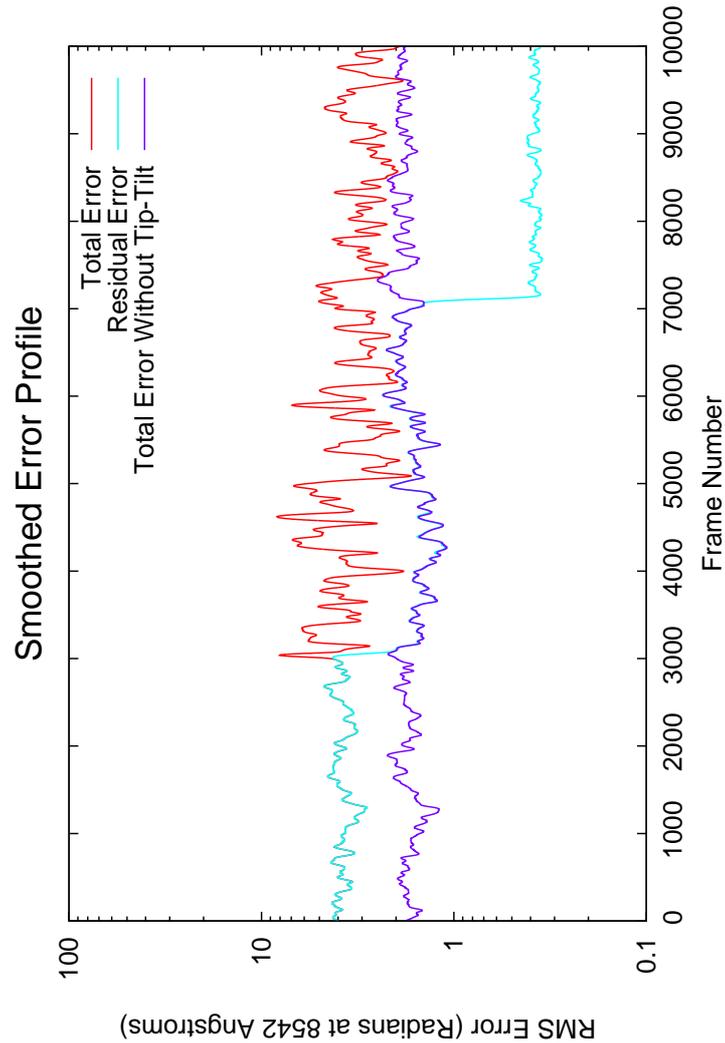}
\caption{Wavefront error in radians at 8542 {\Ang}. These data have been 
smoothed by convolving them with a Gaussian kernel, with a standard 
deviation of 15. The total error calculation becomes very noisy, once 
the TTM is activated, but the total, subtracting tip-tilt errors stays well 
behaved, even when the AO is fully activated. Since the reconstructed tip-tilt 
errors were not used for any calculations, this 
is of little consequence.}
\label{fig:er1}
\end{figure}

\begin{figure}[t!]
\centering
\includegraphics[width=0.8\textwidth,clip]{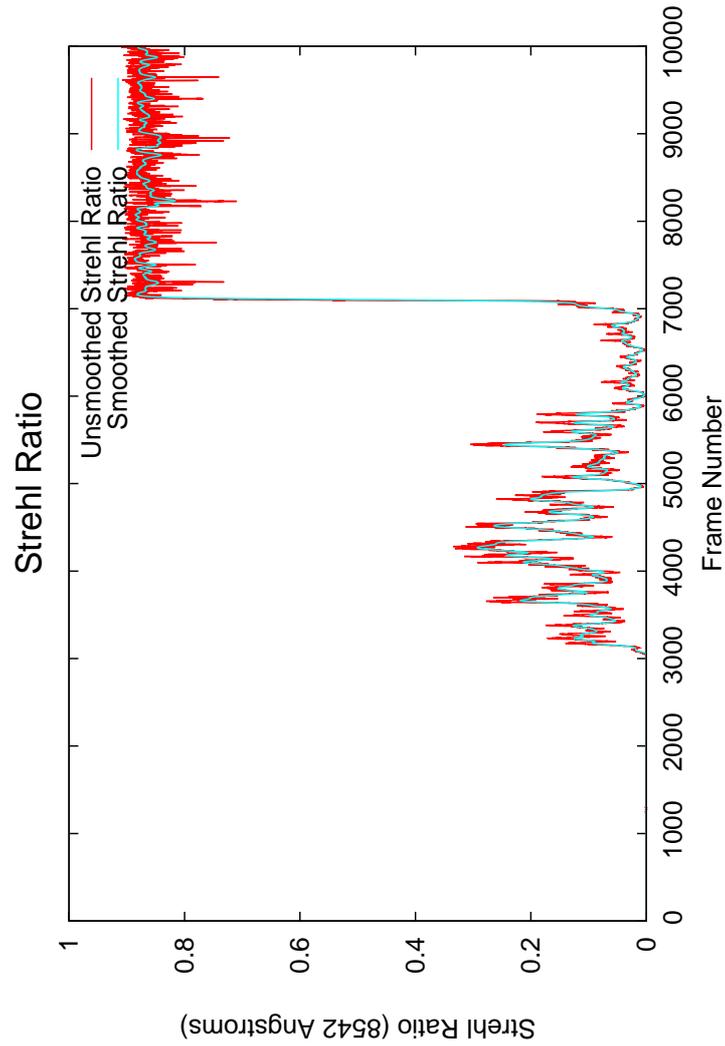}
\caption{Strehl ratio at 8542 {\Ang} for the data plotted in Figure 
\ref{fig:er1}. These data have been smoothed by 
convolving them with a Gaussian kernel,
with a standard deviation of 15. Although the Strehl ratio is at times higher 
than 0.2, when only the TTM is active, it is highly variable.}
\label{fig:st1}
\end{figure}

In Figure \ref{fig:er1}, the improvement in image quality given by 
the off-limb solar AO system is shown. For the first 3000 frames, the AO is 
applying no correction, the residual error is the same as the total 
atmospheric error, as measured by KAOS. At around frame 3000, the system 
begins correcting for 
tip-tilt errors only. The residual error becomes the same as the total 
error minus 
the tip-tilt errors. Finally, at around frame 7000, full AO 
correction is applied, the residual error drops far below the total error 
lines. The seeing was quite good with, $r_0$ of around 17cm.

Figure \ref{fig:st1} Shows the corresponding Strehl ratios, calculated 
during the same sequence as that in Figure \ref{fig:er1}. Although the 
seeing is good, the improvement allowed by correcting for tip-tilt 
errors alone is highly variable and much lower than that given by making the 
full correction.

\section{Conclusions}

The performance of the new off-limb solar AO system has been demonstrated. 
This system has been shown to perform well, even under less-than-ideal 
seeing. It performs exceedingly well during good seeing. This is the 
first time to the knowledge of the authors that any group has used the light of 
solar prominences directly to measure the wavefront in an AO system. 
This allows for superior correction of solar prominence images, in 
the near infrared, when compared to what was previously available.

The final Strehl ratio obtained by this system has been shown to exceed 
the predictions which was made. The Strehl ratio is high enough,
under most circumstances, to allow for the reconstruction of fully 
diffraction-limited images, at 8542 {\Ang}. This allows for spectroscopy and 
spectro-polarimetry at the diffraction-limit of the telescope.

With the significant improvements allowed by the off-limb solar AO system, 
solar physicists will be able to further probe the mysteries of solar 
prominences, by deducing their properties at smaller spatial scales than 
previously possible. The authors believe that this will greatly improve 
humankind's understanding of these phenomena and accompanying processes, 
such as CMEs.


%

%

%

%
%

%
%
 \bibliographystyle{spr-mp-sola}
 \bibliography{AstroRefs}  
%
%
%
%

\end{article} 
\end{document}